\documentclass[prb,aps,twocolumn,showpacs,preprintnumbers, superscriptaddress, amsmath,amssymb]{revtex4-1}
\usepackage{comment}
\usepackage{xcolor} 
\usepackage{soul}
\usepackage{microtype}
\usepackage{amsmath}
\usepackage{dcolumn}
\usepackage{amssymb}
\usepackage{amsthm}
\usepackage{mathrsfs}
\usepackage{graphicx}
\usepackage{verbatim}
\usepackage{bm}
\usepackage{bm}
\usepackage{times}
\usepackage{lipsum}

\usepackage[all]{xy}

\newcommand{\nn}{\nonumber}

\newcommand{\dg}{\dagger}

\newcommand{\ve} {\varepsilon}

\newcommand{\be}{\begin{eqnarray}}
\newcommand{\ee}{\end{eqnarray}}
\newcommand{\la}{\langle}
\newcommand{\ra}{\rangle}
\newcommand{\rar}{\rightarrow}

\begin{document}
	\title{Universal scaling of quantum state transport in one-dimensional topological chain under nonadiabatic dynamics}
	\author{Lingzi Huang}
	\altaffiliation[]{These authors contributed equally to this work.}
	\author{Menghua Deng}
	\altaffiliation[]{These authors contributed equally to this work.}
	\author{Chen Sun}
	\email{chensun@hnu.edu.cn}
	\author{Fuxiang Li}
	\email{fuxiangli@hnu.edu.cn}
	\affiliation{School of Physics and Electronics, Hunan University, Changsha 410082, China}
	
	\begin{abstract}
		When a system is driven across a continuous phase transition, the density of topological defects demonstrates a power-law scaling behavior versus the quenching rate, as predicted by Kibble-Zurek mechanism. In this study, we generalized this idea and  address the scaling of quantum state transport in a one-dimensional topological system subject to a linear drive through its topological quantum phase transition point. We  illustrate the power-law dependencies of the quantum state's transport distance, width, and peak magnitude on the driving velocity. Crucially, the power-law  exponents are distinct for the edge state and bulk state. Our results offer a novel perspective on quantum state transfer and enriches the field of Kibble-Zurek behaviors and nonadiabatic quantum dynamics. 	\end{abstract}
	\maketitle

\section{Introduction}
Quantum state transfer in quantum networks is of crucial importance in quantum control and large-scale quantum information processing \cite{Bennett1992, Bennett1993}. To achieve efficient quantum state transfer across quantum networks, various transfer mechanisms have been proposed, leveraging different physical systems.  Among these mechanisms, most rely on one-dimensional quantum chains with either static or dynamic parameters \cite{Bose2003, Christandl2004}. These schemes can be realized in a variety of physical systems including photonic lattice \cite{Bellec2012,Perez2013,Chapman2016}
, acoustic system \cite{Shen2019}, nitrogen-vacancy centers in diamond \cite{Yao2011},  superconducting qubit circuits \cite{Mei2018, Tsomokos2010}, chains of tunnel coupled quantum dots \cite{Petrosyan2006}, driven optical lattices \cite{Chen2011}, NMR \cite{Cappellaro2007}, and so on.

The efficiency of quantum state transfer schemes is predominantly dictated by two main factors: the transfer speed and the fidelity \cite{Ashhab2012, Caneva2009, Deffner2017, Yung2006, Zhang2018}. Often, these factors are at odds, presenting a trade-off between high speed and high fidelity. While adiabatic quantum evolution can ensure perfect state transfer, it typically results in slow transfer speeds. Conversely, pursuing high transfer speeds usually incurs non-adiabatic excitations, compromising the fidelity of the quantum state\cite{Gras, Knysh,Barends}. To reconcile the two conflicting factors, the emerging field of topological states of matter provides a promising platform \cite{Hasan2010, Qi2011,Sarma2006,Gardas}.  For instance, recent studies of high-fidelity quantum state transfer have utilized the Su-Schrieffer-Heeger (SSH) model to realize robust and fast quantum state transfer protocols \cite{Estarellas, Lang, Longhi2019, Longhi20192,DAngelis2020,Xu2023}. The SSH model is particularly advantageous as it features edge states that are inherently resistant to external disorder, owing to topological protection  \cite{Kitaev,Asboth}.

Most quantum state transfer protocols operate under adiabatic conditions, typically requiring an instantaneous energy gap to achieve high fidelity. A prominent example of this is Thouless pumping, which adiabatically transfers a quantized number of electrons over a periodic cycle\cite{Thouless1983}. It poses an intriguing question: what occurs when the system is nonadiabatically driven across a quantum critical point where the gap closes \cite{Gras, Knysh,Barends}. When a system is quenched across a continuous phase transition,  topological excitations are formed and the density of these excitations demonstrates a universal power-law scaling relationship with the quenching speed. This phenomena, known as Kibble-Zurek mechanism (KZM) \cite{Kibble1976, Kibble1980, Zurek1985, Zurek1996,Lee2015}, has been experimentally verified in a variety of  platforms \cite{Deutschlander, Maegochi, Du2023, Weiler, Lamporesi, Navon, Anquez, Ko2019, Yi2020, Keesling, Ebadi}.  Recent research has expanded on the KZM, uncovering that universal characteristics are also present in rapid quench regime, in full counting statistics of defects and significant fluctuations, thereby enhancing the understanding of nonequilibrium and nonadiabatic dynamics \cite{Campo2018, Zeng2023, Balducci2023}.  This raises an interesting question: are there more physical quantities that exhibit a scaling relationship with the quenching rate if the system traverses a critical point with a closing gap?

In this paper, we investigate the scaling behavior of quantum state transport in a one-dimensional topological chain  when the system is linearly driven across the topological quantum phase transition point. We focus on the transport distance, the width and the peak magnitude of the quantum state, revealing that each of the quantities exhibits a power-law scaling with the driving speed. Crucially, the power-law scaling exponents display distinct values for the edge state and bulk state. We establish that our findings are  applicable not only to the Hermitian SSH model, but also to other 1D topological system such as Creutz ladder model \cite{Creutz ladder,Jafari2019} and non-Hermitian SSH model. Our research offers a novel insight for the quantum state transfer and contributes to the broader understanding of Kibble-Zurek behaviors and nonadiabatic quantum dynamics.

This paper is organized as follows. In Section \ref{sec2}, we introduce the quenching protocol in an SSH model. In Section \ref{sec3}, the scaling behaviors of travel distance, the width and the peak magnitude are discussed. In Section \ref{sec4}, we present theoretical arguments on the scaling behaviors. Sections  \ref{sec5} and  \ref{sec6} are on results on Creutz ladder model and non-Hermitian SSH model, respectively. Finally, discussions and conclusions are presented in Section \ref{sec7}. Appendices A and B are on calculations on probabilities projected to each extended eigenstate for linear quenches from an edge and on discussions of fidelity of adiabatic transfer, respectively.


\section{SSH model and quenching protocol}\label{sec2}
We start with the 1D SSH model to study the quantum state transport during a nonadiabatic dynamics when the system is driven across the topological phase transition point. Our results are also applicable to other 1D topological system, such as Creutz ladder model and non-Hermitian SSH model, as presented in later sections. The Hamiltonian of an open-boundary SSH chain with $N$ unit cells  reads
\be
H=\sum_{n=1}^N J_1 a_{n}^{\dg} b_{n} +\sum_{n=1}^{N-1}  J_2  a_{n+1}^{\dg} b_{n} + h.c.,
\ee
where $a^{\dg}_{n}$ ($a_{n}$) and $b^{\dg}_{n}$ ($b_{n}$) are the creation (annihilation) operators on $A$ and $B$  sublattice in $n$-th  unit cells. $J_1$ and $J_2$ are intracell and intercell couplings, respectively,  and $h.c.$ denotes Hermitian conjugate of all previous terms.  It has been very well understood that the system undergoes a phase transition from topologically nontrivial phase ($J_1<J_2$) to trivial phase ($J_1>J_2$) with gap closing at $J_1=J_2$.  

We consider a quenching process that an initial state $|\psi(t_i)\ra$ 
evolves under the Shr\"{o}dinger equation $i\frac{d}{dt}|\psi\ra=H(t)|\psi\ra $ to a final state $|\psi(t_f)\ra$ with a time-dependent Hamiltonian $H(t)$. We consider two different quench protocols.  The first one is called ``linear quench'' with the couplings varying with time:
\be
J_1=\beta t,~~J_2=1-\beta t.  \label{eq:J1J2}
\ee
 Quench is taken from $t_i=0$ to $t_f=1/\beta$, with $\beta>0$ characterizing the speed of the quench. Such a quench connects the two fully dimerized limits $J_1=0$, $J_2=1$ and $J_1=1$, $J_2=0$. The other is called ``periodic quench'' which consists of two successive linear quenches, the quench \eqref{eq:J1J2} and its mirror reflection about the $t=1/\beta$ line, 
such that the system  returns to the original Hamiltonian after one period, and can thus be comparable to a Thouless pumping process if quenching rate $\beta$ is sufficiently small.

 We will consider two different initial states:
 \be
 |\psi(t_i)\ra=|1,A\ra \label{eq:psi_edge}
 \ee
 which is located at the left edge, and
 \be
 |\psi(t_i)\ra=(|n,B\ra+|n+1,A\ra)/\sqrt 2 \label{eq:psi_m}
 \ee
 which is located in the bulk of the chain, and is far away from the edge.  For example, we can choose $n=N/2$. The two initial states are both eigenstates of the initial Hamiltonian $H(t_i)$.

\section{Scaling of travel distances} \label{sec3}
\begin{figure}[t]
		\centering
		\includegraphics[width=0.48\textwidth]{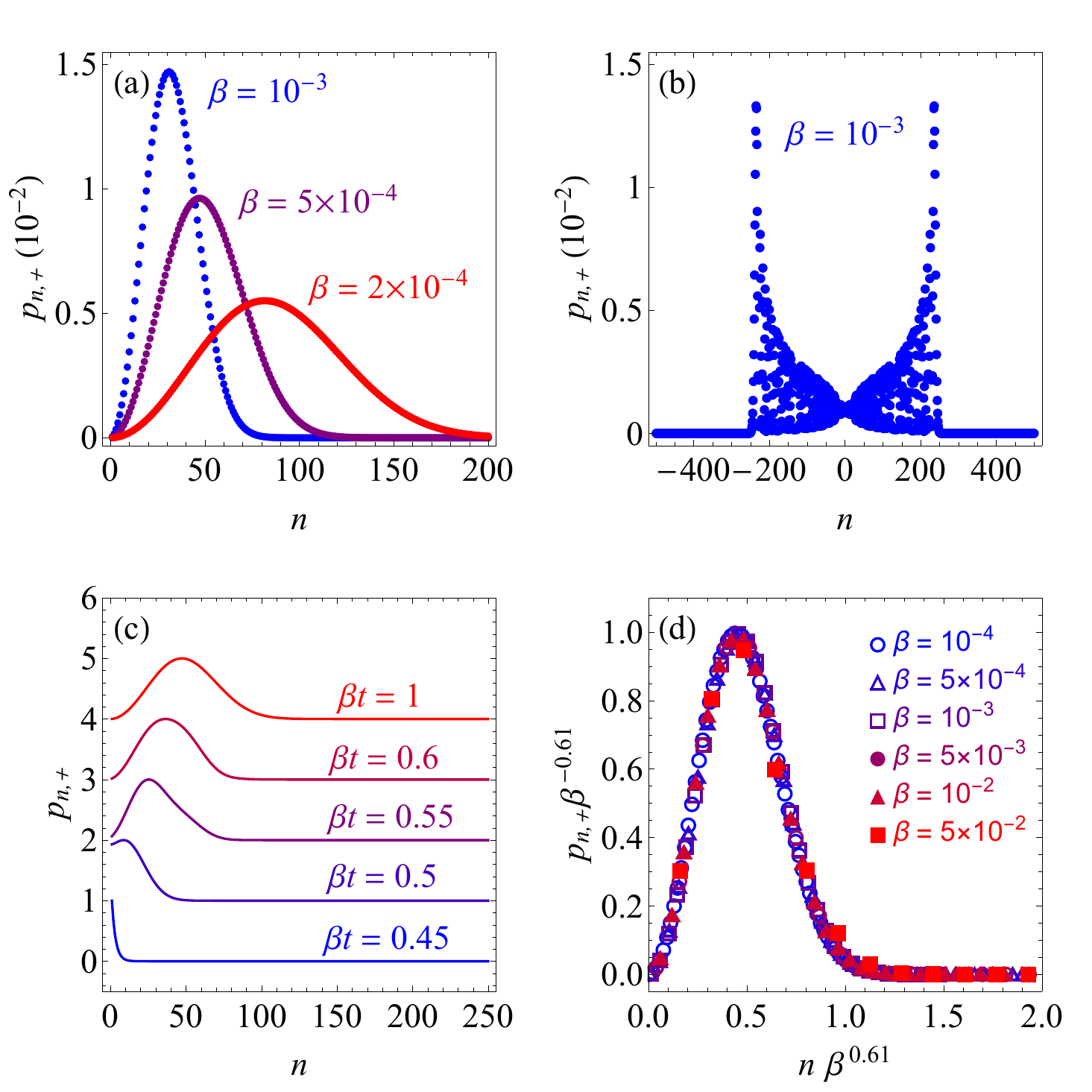}\\
		\caption{Distribution of probabilities $p_{n,+}$ vs. $n$ for  different initial states. (a) and (b) are the distributions of final probabilities for initial edge state (\ref{eq:psi_edge}) and for bulk state (\ref{eq:psi_m}), respectively. In (b) the $n$ axis is shifted such that the initial bulk state is at $n=0$.  (c) Probability profiles $p_{n, +}$ at different times for a fixed quenching rate $\beta=5\times 10^{-4}$. Each profile is stretched vertically so the maximum $p_{n, +}$ becomes $1$.  (d) Probability profiles $p_{n, +}$ of initial edge state for different values of $\beta$ collapse to a single curve after scaling of $n$ and $p_{n, +}$.
}\label{fig1}
	\end{figure}

We first consider the linear quench (\ref{eq:J1J2}). In the adiabatic limit $\beta\rar 0$, since the edge state is always gapped (even though small) from other states, $|1,A\ra$ evolves adiabatically to the right edge state $|N,A\ra$. 
In the opposite diabatic 
limit $\beta\rar \infty$, $|1,A\ra$ remains unchanged simply because there is no time to evolve. We are interested in the intermediate region between these two limits, where the final state is expected to be distributed over the chain. 
By numerically solving the time-dependent Schr\"{o}dinger equation, we calculated $|\psi(t_f)\ra$ for an open SSH chain with $N=1000$ unit cells from an initial state at the left edge (Eq.~\ref{eq:psi_edge}) or at the middle of the chain (Eq.~\ref{eq:psi_m}). To characterize the final state, we define the final probabilities
\be
p_{n,\pm}=|\la n,\pm|\psi(t_f)\ra|^2,
\ee
 which are the probabilities of final state projected to final localized eigenstates $|n,\pm\ra =(|n,A\ra \pm |n,B\ra)/\sqrt2$.  Note that at final time $t_f$, the Hamiltonian is dimerized, and $|n, \pm\ra$ are its eigenstates with energies being $\pm 1$. 
Calculations show that for $|\psi(t_i)\ra=|1,A\ra$ we always have $p_{n,+}=p_{n,-}$, which is due to chiral symmetry of the system. For bulk state (\ref{eq:psi_m}) with $n=500$,  $p_{n,-}$ is much smaller than $p_{n,+}$  due to the fact that the initial state $|\psi(t_i)\ra$ is an eigenstate of the initial Hamiltonian with positive eigenvalue. Thus, it suffices to focus on $p_{n,+}$. One sees qualitatively different behaviors for the two types of initial states: for evolutions from the edge (Fig.~\ref{fig1}(a)), $p_{n,+}$ shows a smooth profile with a single peak, resembling
a coherent state; whereas for evolutions from the middle (Fig.~\ref{fig1}(b)), $p_{n,+}$ develops oscillations whose strength increases when moving away from the initial position, until a peak emerges followed by a sharp fall-off to zero at large $n$.  We also note that, as depicted in Fig.~\ref{fig1}(c), the transport doesn't start until near the critical point at time $t=0.5/\beta$. This observation is helpful in understanding the scaling behaviors later.

\begin{figure}[t]
		\centering
		\includegraphics[width=0.5\textwidth]{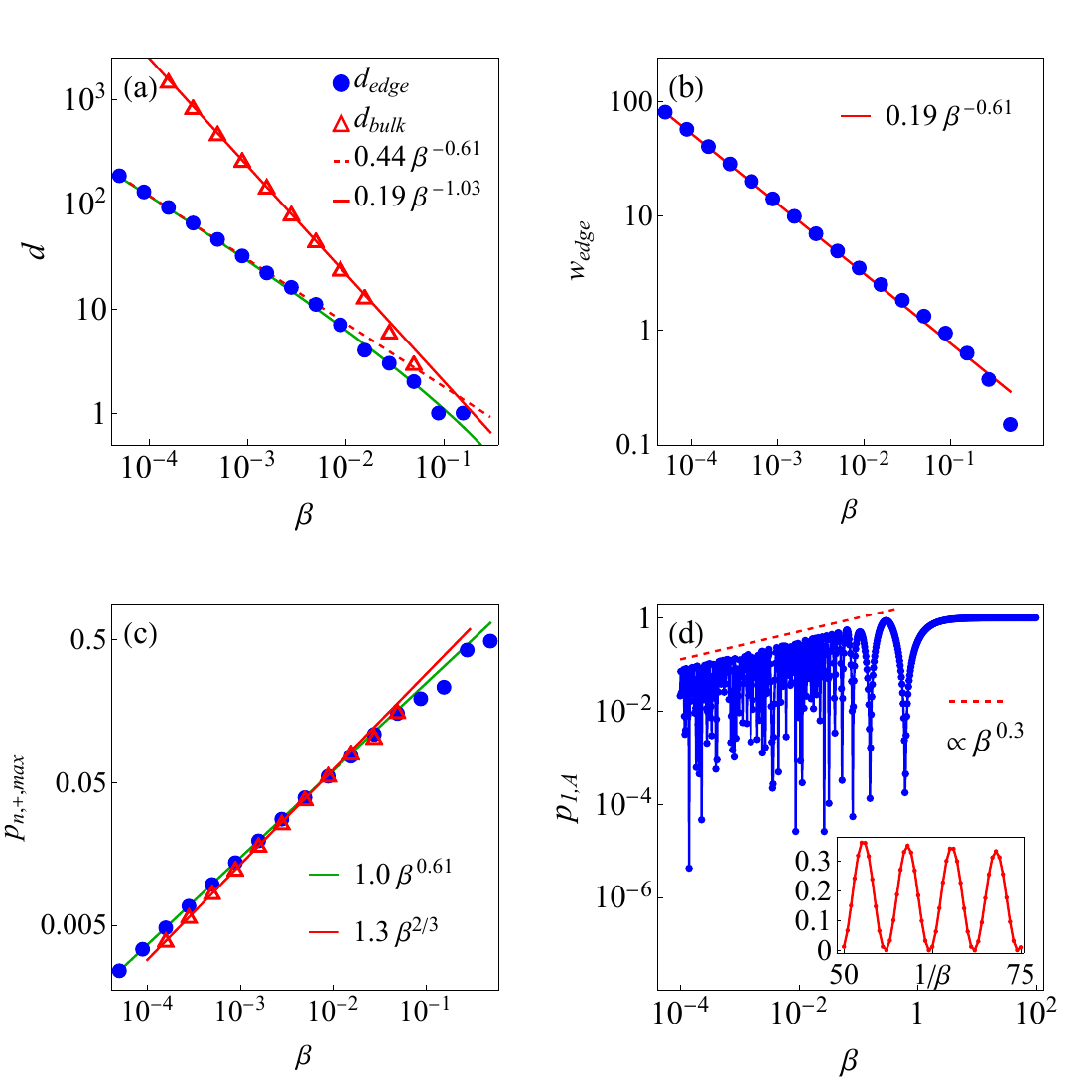}\\
		\caption{Scalings in the open SSH chain for $N=1000$:  (a) scaling of travel distance, (b) scaling of the width, (c) scaling of $p_{n,+}$ at the peak.   Numerical results are represented by plot markers, blue dots for initial edge and red triangles for initial bulk state. Solid lines are power-law fittings.  
		(d) Probability $p_{1,A}$ to return to the edge for a periodic quench from an initial edge state. The inset shows that $p_{1,A}$ oscillates with a period $2\pi/\beta$.  }\label{fig2}
	\end{figure}

The scaling behaviors of the initial edge state can be summarized in Fig.~\ref{fig1}(d), in which, the probability distribution $p_{n, +}$ for different values of quenching rate $\beta$ collapse to a single curve after one rescales $n$ and $p_{n, +}$  by $\beta$. This indicates that all the quantities, such as travel distance and peak magnitude, that are related to the transported quantum state would exhibit power-law relations with quenching rate, as illustrated in Fig.~\ref{fig2}.
First, we consider how far the state travels under the quench.
We define the position of the peak as the traveled distance $d$. In Fig.~\ref{fig2}(a), $d$ exhibits  a power-law scaling for both initial edge state and bulk state $d\propto \beta^{-\nu}$, but with different scaling exponents.  For edge state, $\nu\approx 0.61$, while for bulk state, $\nu\approx 1.03$.
For initial edge state, one can also study the width of the profile of $p_{n,+}$, defined by the standard variance
$w_{edge}=[ \sum_n (n-d)^2 p_{n,+}]^{1/2}$.
As plotted in  Fig.~\ref{fig2}(b), it also exhibits a power-law scaling with the same exponent as the travel distance:
$w_{edge} \propto\beta^{-0.61}$.

Moreover, in Fig.~\ref{fig2}(c)  the maxima of probability distribution $p_{n,+}$ of both initial edge state and initial bulk state are plotted as a function of quenching rate $\beta$. One sees that again the maximum exhibits a power-law scaling with $\beta$, but with different scaling exponents for the two types of initial states. For initial edge state, the exponent is roughly $0.61$, while for initial bulk state, $2/3$ can fit the numerical results very well. 

We note that in all our calculations $p_{n,+}$ decay fast to zero for $n$ sufficiently far away from the initial positions. Under periodic boundary conditions, we find that the results for evolutions from the bulk also fulfills the same scalings, $d_{PBC}\propto \beta^{-1.03}$ and  $p_{n,+,max}\propto \beta^{2/3}$. 
We note that different boundary conditions or different length of lattice don't change  these results as long as the end of the chain is not reached by the state evolution. 



For periodic quench, after one periodic circle, one would expect that the state would transfer from one edge state to another edge state located at the other end of the chain, if the quench is slow  enough so that the system is under adiabatic regime, which is called Thouless pumping. However, when crossing a topological phase transition point with gap closing, near this gap-closing point the adiabatic condition always breaks no matter how slow the quench is; this follows from the adiabatic theorem which states that conditions of an adiabatic evolution are that the quench is slow enough and that the spectrum of the system is always gapped. Thus, the adiabatic condition cannot be fully satisfied, leading to nonzero populations to excited states, and thus the quantized state transfer cannot be realized. Here in order to quantify the state transfer, we examine the final probability at the first lattice point $p_{1, A}$ after one periodic quench.
Plotting $p_{1,A}$ vs. $\beta$ for quenches from an edge in Fig.~\ref{fig2}(d), we see that $p_{1,A}$ oscillates between zero and an upper bound $p_{1,A,max}$ which scales as $\beta^{0.3}$. Plotting $p_{1,A}$ vs. $1/\beta$ shows that this oscillation has a fixed period $2\pi/\beta$ (inset of Fig.~\ref{fig2}(d)). 



It's interesting to see whether the observed scaling laws apply to other quench protocols. For evolution from an edge, we considered a quench with a sinusoidal shape,
\be
J_1=\sin^2(\beta t), ~~J_2=\cos^2(\beta t),
\ee
from $t_i=0$ to $t_f=\pi/(2\beta)$. We find that the behavior of $p_{n,+}$ is similar to a linear quench shown in Fig.~\ref{fig1}.

For evolution from the bulk, we considered the above sinusoidal quench and a sudden quench where $J_1$ and $J_2$ are both set to non-zero constants from $t=0$ to $t_f=1/\beta$, and find the same scaling laws $d_{bulk}\propto \beta^{-1}$ and $p_{n,+,max}\propto \beta^{2/3}$. 

\section{Theory of quenches}\label{sec4}
In this section, we present theoretical arguments on the scaling behaviors of the quantum state transport under nonadiabatic quench dynamics discussed above.

\subsection{Theory of quench from an edge}
To understand the scaling behaviors, we first look at the spectrum of the system. For a chain with odd number of sites, the eigenvalues of the Hamiltonian have analytical forms \cite{Coutant_2020}: there is a single edge state with  zero energy, and $2N-2$ extended states with energies
$$ E=\pm\epsilon_j=\pm|J_2+J_1e^{i k_j}|$$ where $k_j=\pi j/N$ and $j=1,2,\ldots ,N-1$. 
As a result of such a spectrum, during an evolution the edge state is gapped from all the extended states except near time $t=t_f/2=1/(2\beta)$. For an evolution starting from the left edge,  we expect that during the quench the state should remain near the left edge until close to $t_f/2$ when the gaps become minimal, see Fig.~\ref{fig1}(c). At $t_f/2$, a portion of the state transfers to each extended state, which then travels at the group velocity of the extended state
\be
v_k=d\epsilon_j/dk=J_1J_2\sin( k_j)/\epsilon_j,
\ee
 and the  whole state is a superposition of traveling of each extended state. 
In Appendix A, we further justify this picture by calculating probabilities projected to each extended eigenstate.

Based on this picture, we can estimate the travel distance $d_{edge}$. Note that the Hamiltonian depends linearly on time and thus constitutes a multi-level Landau-Zener (LZ) problem, which may not be solved exactly. The traditional LZ problem is to find the transition probabilities from an initial state to each
final energy state. The exactly solvable models are rare and have been found to exist only in some special forms \cite{Sinitsyn2014, Sinitsyn2016, Li2017, Li2018, LZSM}. Here, instead of looking for an exact solution, we provide an Ansatz that can be fitted by numerical results. We observe that the transition probability   to each final state (denoted as $p_k$ \cite{note-pk})  is solely determined by a dimensionless parameter $\Delta_k^2/\beta$, where $\Delta_k=\epsilon_k=\cos(k/2)$ is the gap between an extended state and the edge state at $t=t_f/2$. One can use the following Ansatz:
\begin{align}\label{p_vs_epsilon_fit}
p_k=c_3 \left(1 - e^{-\frac{ c_1 \Delta_{k}^2}{\beta}}\right)e^{-\frac{ c_2 \Delta_{k}^2}{\beta}}.
\end{align}
$c_1$ and $c_2$ are two fitting parameters, and $c_3$ is an overall normalization factor. Comparison with numerical results are presented in Appendix A. Note that $p_k$ is small both at small and at large $\Delta_k$, and it reaches a peak at $\Delta_k^2/\beta\approx 1$.
The wavevector corresponding to this peak is $k=2\arccos\sqrt{\beta}$, which can be used to estimate the travel distance as
\begin{align}\label{}
d_{edge}=\int_{1/(2\beta)}^{1/\beta}|v_k| dt
\end{align}
assuming that the traveling roughly starts at time $1/(2\beta)$ and ends at time  $1/\beta$. This integral can be analytically performed, giving
\begin{align}\label{eq:arctan-fit}
&d_{edge}=\frac{1}{8\sqrt \beta}  \left(\frac{2-  \beta }{1-  \beta } \operatorname{arccosh}\frac{1}{ \sqrt{ \beta }}-\sqrt{\frac{1}{1-  \beta }}\right).
\end{align}
The green curve in Fig.~\ref{fig2}(a) is Eq.~\eqref{eq:arctan-fit}, which agrees well with the exact $d_{edge}$, and at large $\beta$ it correctly predicts the drop of $d_{edge}$ below the power-law scaling.

The power-law scaling of $d_{edge}$ can now be understood. Since $ \beta\ll 1$ in almost the whole range of $\beta$ considered, keeping the leading order in $ \beta$ gives
\begin{align}\label{}
&d_{edge}\approx\frac{1}{8\sqrt{\beta}}  \left(\log\frac{4}{ \beta}-1\right).
\end{align}
Thus, $d_{edge}$ scales as $d_{edge}\propto \beta^{-1/2} \log(1/\beta)$, which is visually close to a power law $d_{edge}\propto \beta^{-\nu}$ with some $\nu>1/2$ within a range of $\beta$.

\subsection{Theory for quench from the bulk} 
As mentioned before, for evolution from the bulk we could consider a chain with periodic boundaries. Now the Hamiltonian can be reduced to a $2\times 2$ matrix in the momentum space labeled by $k=2\pi j/N$ ($j=0,1,\ldots ,N-1$), with eigenenergies $E=\pm\epsilon_k=\pm|J_1+J_2e^{-i k}|$. Let's denote the corresponding eigenvectors in the $k$th-block as $|\psi_{k,\pm}\ra$. The initial localized state $|\psi(t_i)\ra=(|n_i,B\ra+|n_i+1,A\ra)/\sqrt 2$ can be written as a superposition of 
all eigenstates with positive energies:
\begin{align}\label{}
|\psi(t_i)\ra=(1/\sqrt N)\sum_k e^{-i k n_i}|k\ra\otimes |\psi_{k,+}\ra.
\end{align}
During the quench each $|k\ra\otimes |\psi_{k,+}\ra$ travels with velocity $v_k=\partial\epsilon_j/\partial k$, and accumulates to a distance $\int_0^{1/\beta}v_k dt$ at the final time \cite{note-LZ}. Note that unlike evolution from the edge, the traveling starts immediately at $t=0$.
We can estimate the travel distance $d_{bulk}$ as the largest distance traveled by all the eigenstates, which simply gives rise to  $d_{bulk}=c/\beta$ obtained by a change of variable $t\rar t/\beta $. Here, $c$ is a constant and numerically determined to be around $0.24$.


For the understanding of the magnitude of peak $p_{n,+,max}$, we consider the corresponding amplitude  that can be approximated by assuming adiabatic evolution
\cite{note-adiabatic}:
\be
\la n_{max},+|\psi(t_f)\ra=(1/ N)\sum_k e^{-i k n_i}e^{i( k n_{max}-\varphi_k )},
\ee
 where $\varphi_k=\int_{0}^{1/\beta}\epsilon_k dt $ is the adiabatic phase accumulated during the evolution of $|\psi_{k,+}\ra$. The magnitude can be estimated by the stationary phase approximation. Let's consider Taylor  expansion of $\varphi_k$ near $k_{max}$, which is the momentum labeling each final state with maximum transition probability. Therefore, one has $(\partial^2\varphi_k/\partial k^2)_{k_{max}}=0$.
Since $\partial\varphi_k/\partial k=\int_{0}^{1/\beta}v_{k} t$, we have $(\partial\varphi_k/\partial k)_{k_{max}}=d_{bulk}=n_{max}-n_{i} $.  
Thus, the exponent in $\la n_{max},+|\psi(t_f)\ra$ is dominated by the third order expansion $(k-k_{max})^3$ term. It's easy to see that $\varphi_k\propto  1/\beta$, so $(\partial^3\varphi_k/\partial k^3 )_{k=k_{max}}=C /\beta$, where $C$ is a constant. Replacing the summation $\sum_k$ by integration $ \int dk/(2\pi /N)$, we arrive at
\be
\la n_{max},+|\psi(t_f)\approx\frac{1}{2\pi}\int_{-\infty}^{\infty} dk e^{-(i\frac{C}{6\beta})(k-k_{max})^3},
\ee
 which scales as $\propto  \beta^{1/3}$, 
and thus $p_{n,+,max}=|\la n_{max},+|\psi(t_f)\ra|^2\propto  \beta^{2/3}$.


\subsection{Theory of periodic quench}
We now consider the probability $p_{1,A}$ to return to the left edge in a periodic quench, which can be considered as consisting of two linear quenches. The return probability is given by  $p_{1,A}=|\sum_k (A_{k,+}+ A_{k,-})|^2$, where $A_{k,\pm}$ is the amplitude contributed from an extended state labelled by $k$ and $\pm$. It can be written as $A_{k,\pm}=\sqrt{p_{k}}\sqrt{p_{k}'}e^{-i\varphi_{k,\pm}}$, where $p_{k}'$ is the transition probability from the extended state to the edge state during the second linear quench, and $\varphi_{k,\pm}$ is the phase for this amplitude. The sum in $p_{1,A}$ can be simplified by noticing that the multistate LZ model corresponding to the considered linear quench belongs to bipartite models studied in \cite{cross-2017}. According to \cite{cross-2017}, the transition probability matrix of such a model is symmetric, so $p_{k}'=p_{k}$. Moreover, chiral symmetry further implies $\varphi_{k,+}=-\varphi_{k,-}$. 
Thus, we have
\begin{align}\label{eq:p1A}
p_{1,A}=\left(2\sum_k p_{k} \cos\varphi_{k,+}\right)^2.
\end{align}

We may approximate $\varphi_{k,+}$ at $k\sim \pi$ by the adiabatic phase accumulated during the quench, namely, the area under the curve of $\epsilon_k $ in an $E$ vs.$t$ diagram. Since the peak of $p_{k}$ is at $k=2\arccos\sqrt{\beta}$, which for $\beta\ll 1$ is close to $\pi$,  we only need to calculate $\varphi_{k,+}$ for $k\sim \pi$,  which is simply  $\varphi_{k,+}\approx\int_{1/(2\beta)}^{3/(2\beta)}\epsilon_k dt =1/(2\beta)$. Therefore, we arrive at the observed result that $p_{1,A}$ oscillates with a period $2\pi/\beta$. In fact, what takes place here is the phenomenon of Landau-Zener-St\"{u}ckelberg interferometry \cite{LZS-interf-2010}: an interference between the total amplitude of all positive-energy extended states $\sum_k A_{k,+}$ and that of all negative-energy extended states $\sum_k A_{k,-}$ leads to a periodic dependence of the return probability on $1/\beta$. A maximum is reached whenever $\sum_k A_{k,+}$ and $\sum_k A_{k,-}$ are in constructive interference, namely, when $\arg(\sum_k A_{k,+})=0$. Thus, at a given $\beta$ the upper bound of $p_{1,A}$ reads $p_{1,A,max}=\left|2\sum_k p_{k} e^{i\varphi_{k,+}} \right|^2$.
  A calculation of $p_{1,A,max}$ using Eq.~\eqref{eq:p1A} with numerically obtained $p_{k} $ and $\varphi_{k,+}$ agrees well with the upper bound of $p_{1,A}$ in Fig.~\ref{fig2}(d).



\section{Scaling behaviors in Creutz ladder model}\label{sec5}

\begin{figure}[!htb]
\scalebox{0.35}[0.35]{\includegraphics{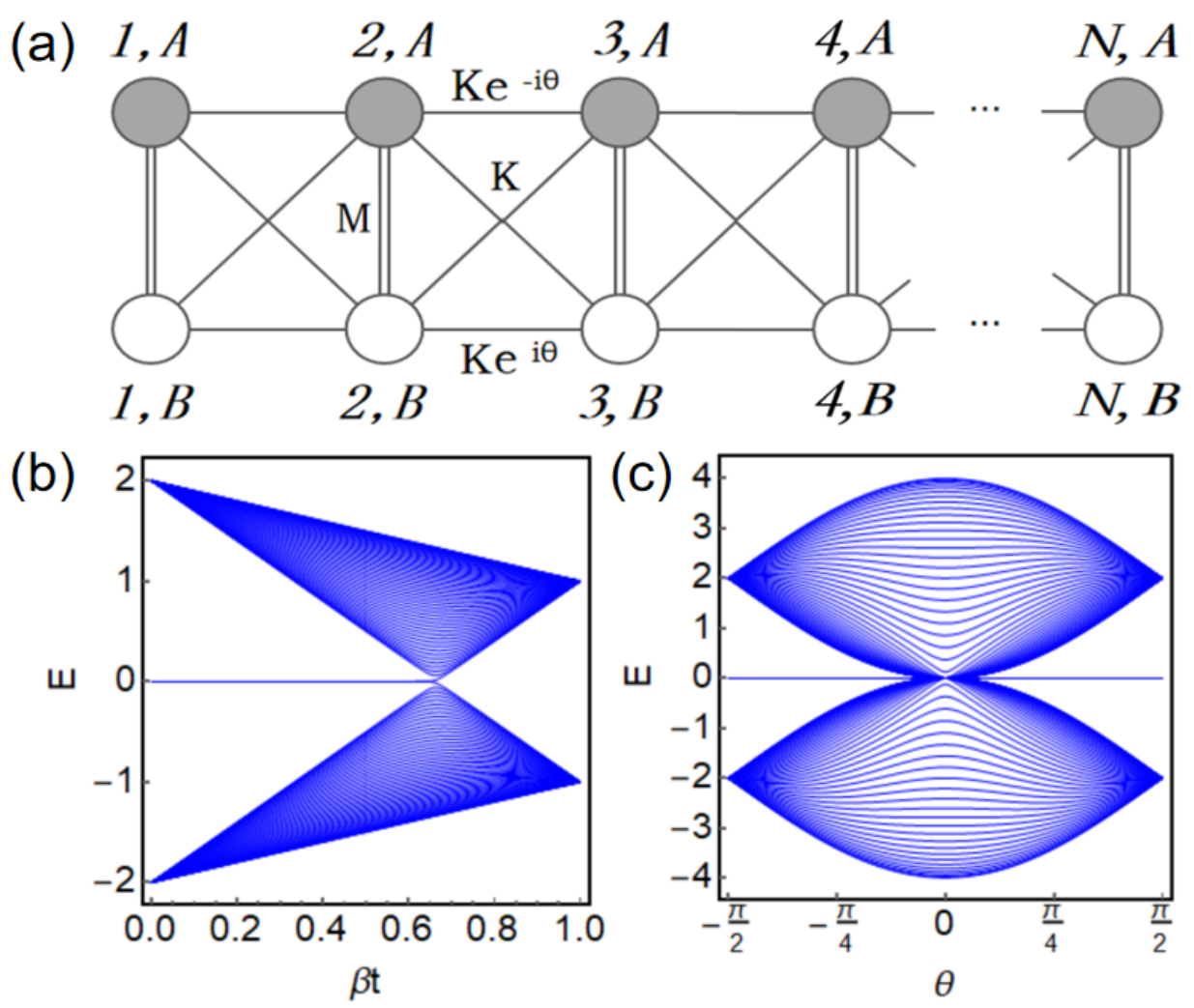}}
\caption{(a) The diagram depicts the structure of the Creutz ladder model with A and B  the two sublattices. $W$, $K$, and $K e^{\pm i \theta}$ are the vertical, horizontal and, diagonal hopping integrals. (b) The spectrum of $E$ vs. $\beta t$ for quenching $M$ and $K$ of the Creutz ladder model. (c) The spectrum of $E$ vs. $\beta t$ for quenching the complex phase $\theta$ of the Creutz ladder model.
}
\label{fig:ladder-model}
\end{figure}

We have considered the SSH model as an example to study the scaling behaviors of quantum state transfer under nonadiabatic quench dynamics. To illustrate that the scaling behavior is not restricted to some special model but is universal, we now consider another model originally proposed by Creutz in \cite{Creutz ladder}. The Creutz model is also a 1D chain consisting of two lattice points $A$ and $B$, as depicted in Fig.~\ref{fig:ladder-model}. The Hamiltonian reads
\begin{align}
&H_{Creutz}=-\sum_{i=1}^{N}  [K(a^{\dg}_n b_{n+1}+b_n^{\dg} a_{n+1})\nn&\\
&\qquad+K(e^{-i \theta} a^{\dg}_n a_{n+1}+e^{i \theta} b_n^{\dg} b_{n+1})\nn\\
&\qquad+M a_n^{\dg} b_{n}+h.c.],
\end{align}
where $n$ label the unit cells with a total of  number of $N$, $A$ and $B$ label the sublattice, $M$, $K$, $Ke^{-i \theta}$ are the vertical, diagonal, and horizontal hopping integrals, respectively. The phase factor $e^{\pm i \theta}$ mimics the presence of a magnetic field which pierces the ladder and supplies a magnetic flux $\theta/\pi$ per plaquette. $a^{\dg}_{n}$ ($a_n$) and $b^{\dg}_{n}$ ($b_n$) are the creation (annihilation) operators on the  sublattice $A$ and $B$ in the $n$-th unit cell, respectively. We consider two different   quenching protocols. The first one is quenching $M$ and $K$
 \be
 M = \beta t, ~~ K = 1 - \beta t,
 \ee
 with fixing $\theta = - \pi /2$ and time changing from $t_i=0$ to $t_f=1/\beta$.
The second one is quenching $\theta$ in the following way
 \be
 \theta = \beta t - \pi / 2\in (-\frac{\pi }{2}, \frac{\pi }{2}),
 \ee  with fixing  $M = 0$, $K = 1$ and time changing   from $t_i=0$  to  $t_f=\pi/\beta$.
 Both of the two quench protocols are described by the quenching rate $\beta$.
The instantaneous eigenvalues of the Hamiltonian as functions of time $t$ are plotted in Fig.~\ref{fig:ladder-model}.

We consider two different initial localized states. One is a plaquette-blocked state
\be
|\psi(t_i)\ra_n=\frac{1}{2} (-ia_{n}^{\dagger}+b_{n}^{\dagger}+a_{n+1}^{\dagger}-ib_{n+1}^{\dagger})|0\ra
\ee localized in the middle of the chain with $n=N/2$.
The other is a topological edge state located at the left edge
\be
|L\ra=\frac{1}{\sqrt{2}} (a_{1}^{\dagger}-ib_{1}^{\dagger})|0\ra.
\ee
Note that these initial states are eigenstates of the initial Hamiltonian.

The two quenching protocols start with the same initial parameters but go through different phases.
Under open boundary conditions,  the initial parameters are $M = 0$, $K = 1$ and $\theta = -\pi /2$, there are two localized chiral zero-energy modes  at the left edge 
and the right edge. 
For the first quench protocol with quenching $M$ and $K$, the system undergoes a phase transition at time $t=0.65/\beta$, after which the system enters into a topologically trivial phase with no edge states(as shown in Fig.~\ref{fig:ladder-model}(b)).  For the second quench protocol with quenching only $\theta$, one can see that there are always two zero-energy modes as shown in Fig.~\ref{fig:ladder-model}(c), even though the system crosses a quantum critical point at $\theta _{c}=0$ with gap closing. 

\begin{figure}[!htb]
\scalebox{0.45}[0.45]{\includegraphics{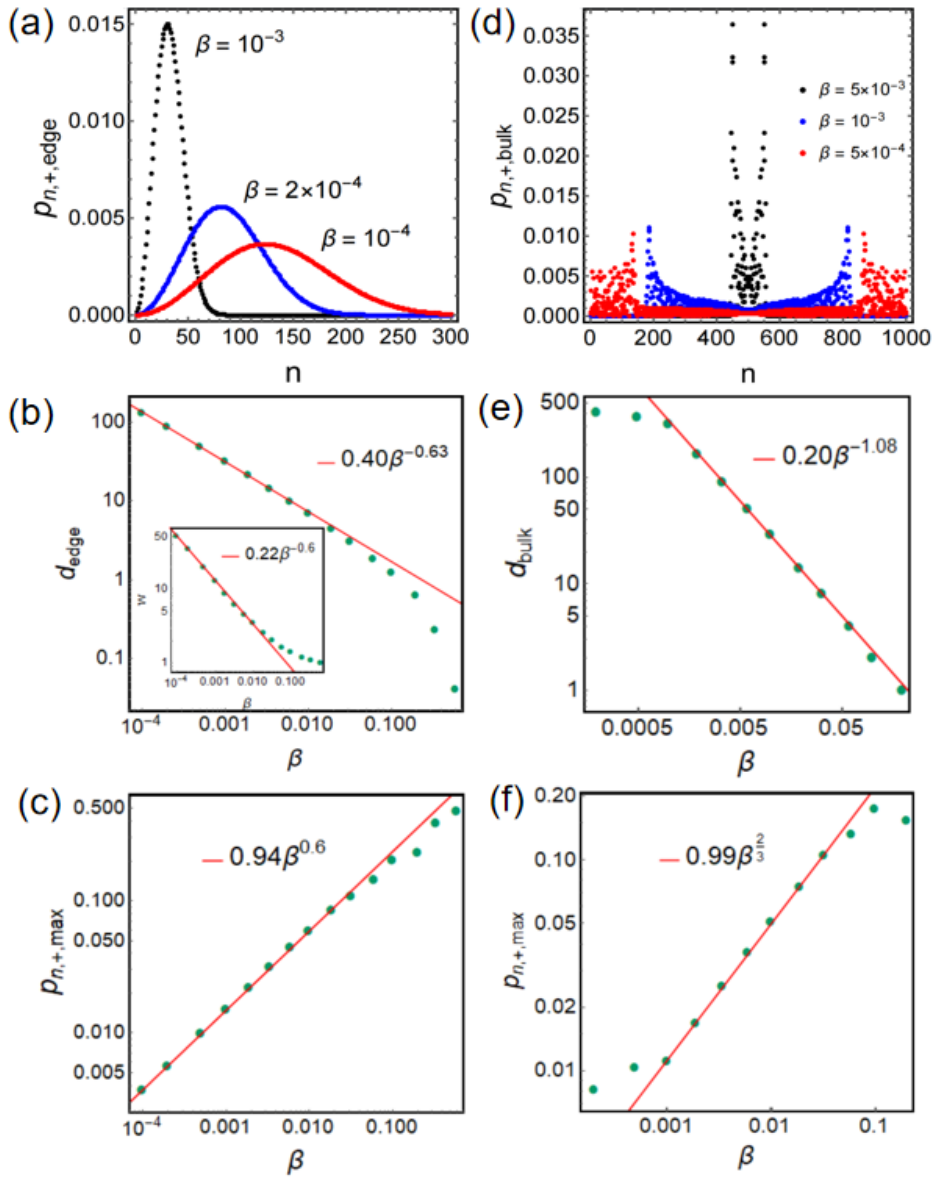}}
\caption{Profile of $p_{n,+}$ and scalings of quenching $M$ and $K$ in the open Creutz ladder for $N=1000$ for different initial states: (a)-(c) left edge state $|L\ra$ (d)-(f) bulk platform $|\psi(t_{i})\ra_n$. (a), (d) are distribution diagrams for two cases at different $\beta$ distinguished by black, blue and red dots. Scalings of travel distances and the maximum of $p_{n,+}$ for $|L\ra$ provided in (b) and (c) shows $d_{edge}\propto \beta^{-0.63}$ and $p_{n,+,max}\propto \beta^{0.6}$. 
(e) and (h) show that bulk platform state transmits at the scaling as $d\propto \beta^{-1.08}$ and $p_{n,+,max}\propto \beta^{\frac{2}{3} }$.
}
\label{fig:scaling ladder}
\end{figure}

We now discuss the scaling behavior of quantum state transfer under quench dynamics of the Creutz ladder model by numerically solving the time-dependent Schr\"{o}dinger equation  for a chain with $N = 1000$. First we study the first kind of quench protocol that quenches $M$ and $K$.  We plot the probability profile $p_n$ of the final state in Fig.~\ref{fig:scaling ladder}(a) for the two different initial states, the left edge state $|L\ra$ and  the plaquette-block state $|\psi(t_i)\ra_{n,+}$ localized in the middle of the chain with $n =500$. Similar to the case of  the SSH model, we can see that the distribution of final states exhibit quite different profiles for the two different initial states. For initial edge state, the final state has a smooth profile for different values of $\beta$, while for initial plaquette-block state, the shape of final state is not smooth, but shows two sharp peaks.

Further, we study the  scaling of travel distances $d$ and the maximum of $p_{n,+}$ as functions of quenching rate $\beta$.  Comparing with the result in SSH model, for quenching from edge state, the travel distance and the maximum of $p_{n,+}$ obey almost the same power-law scaling that $d_{edge}\propto \beta^{-0.63}$ and $p_{n,+,max}\propto \beta^{0.6}$. Additionally, the scalings for evolution from $|\psi(t_{i})\ra_n$ are almost in accordance with that in the case of evolution from SSH middle state, which present $d_{bulk platform}\propto \beta^{-1.08}$ and $p_{n,+,max}\propto \beta^{\frac{3}{2} }$.

 For the second quench protocol, unfortunately, if the initial state is on the edge, e.g. $|L\ra$, the state doesn't move and is always localized on the original location. For the initial plaquette-block state, one obtains the same scaling exponents for the distance and peak of transported wavefunction.

\section{scaling behavior for non-Hermitian SSH model}\label{sec6}

\begin{figure}[!htb]
\scalebox{0.3}[0.3]{\includegraphics{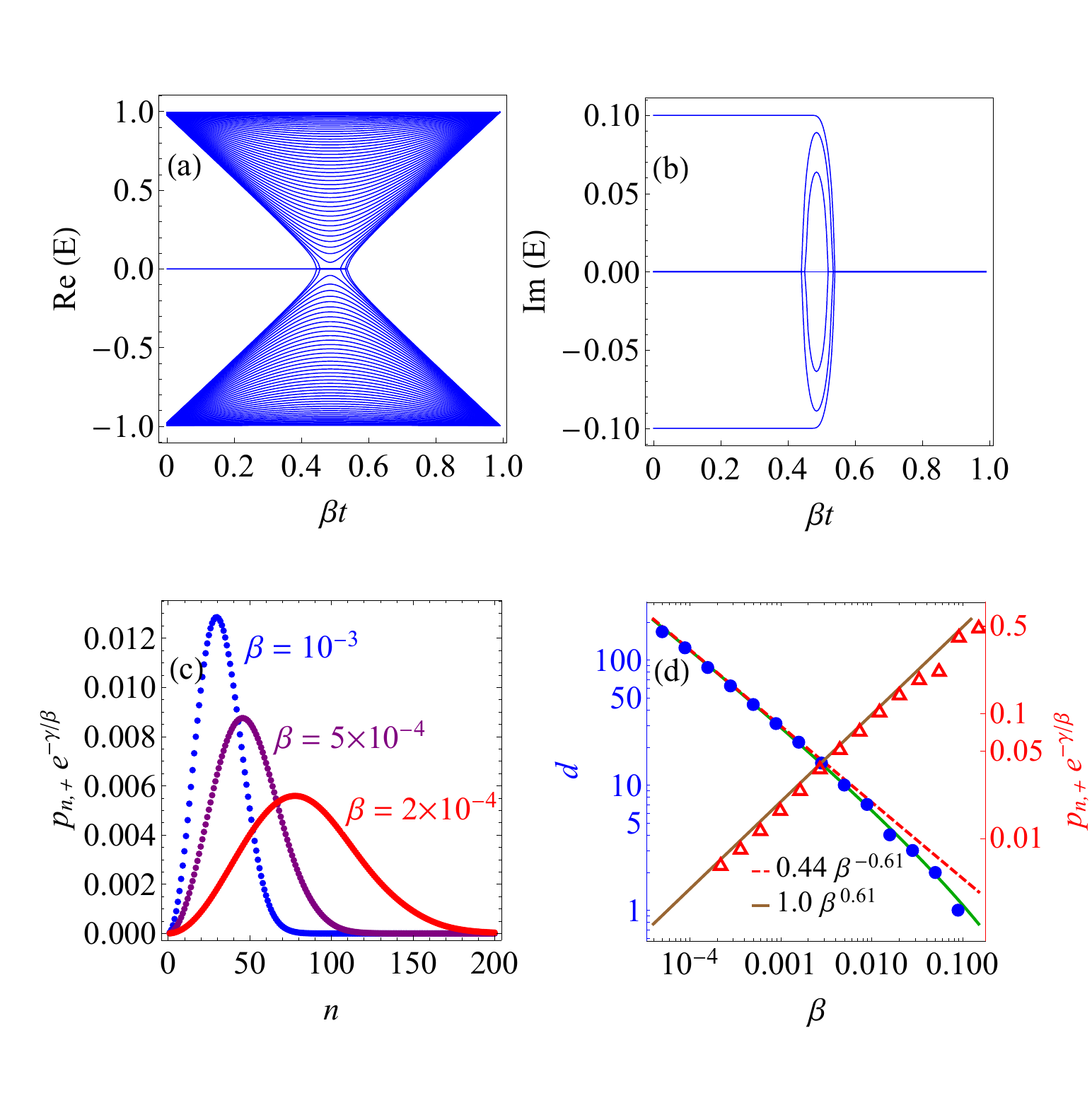}}
\caption{Energy spectrum of a non-Hermitian SSH model and the scaling behavior of travelled quantum state. (a)-(b) The real part and imaginary part of instantaneous energy levels as functions of time.  (c) Rescaled probability profile for different values of quench rate $\beta$. (d) Scalings of travel distances and the maximum of $p_{n,+}$ for initial edge state. {Numbers of unit cells are taken as $N=50$ for (a) and (b), and $N=500$ for (c) and (d).}}
\label{fig:nonHermitian}
\end{figure}

We consider a non-Hermitian SSH model with imaginary on-site energy $i\gamma$ and $-i\gamma$:
\be
H=&&\sum_{n=1}^N [(J_1 a_{n}^{\dg} b_{n} + J_2  a_{n+1}^{\dg} b_{n} + h.c. ) \nn\\
&&+ i\gamma a_{n}^{\dg} a_{n}- i\gamma b_{n}^{\dg} b_{n}].
\ee
Here, we still quench $J_1$ and $J_2$ in such a protocol:
\be
J_1 = \beta t, ~~ J_2 =1-\beta t,
\ee
from time $t_i=0$ to $t_f=1/\beta$, with $\beta$ the quenching rate. Even with the presence of imaginary on-site energy, at initial time, the system is dimerized, and with two edge states located at the two ends of the chain. At final time $t_f$, the system is also dimerized, with eigenenergies: $\ve_{\pm} = \pm \sqrt{1-\gamma^2}$, and eigenstates
\be
|\pm\ra = \frac{1}{\sqrt{2}}[|n,A\ra + (\pm \sqrt{1-\gamma^2} -i\gamma)|n,B\ra ]. \nn \\
\ee
During the quench, the system undergoes two exceptional points at $J_1- J_2 = \pm \gamma$. The instantaneous eigenenergies with open boundary condition are plotted  in Fig.~\ref{fig:nonHermitian}(a) and (b). The probability distributions $p_{n,+}$ for the initial edge state after linear quench dynamics are numerically calculated 
for different values of quench rate $\beta$. 
The profile is similar to the  Hermitian case, but with a very large magnitude of the order of $e^{\gamma/\beta}$ due to the non-Hermitian term.  Nevertheless, after rescaling the probability with a factor $e^{-\gamma/\beta}$, one can still obtain a very well defined probability  profile as shown in  Fig.~\ref{fig:nonHermitian}(c) for different values of $\beta$. The same as in the Hermitian SSH model and Creutz ladder model, we still have scaling behaviors for the travel distance $d$ and peak magnitude $p_{n,+, max}$ with the latter rescaled by $e^{-\gamma/\beta}$. The scaling exponent is the same $0.61$, which indicates a universal value.

\section{Discussions and Conclusions} \label{sec7}
We have considered the dynamical behaviors of  quantum state transport in a 1D topological chain which is driven across the topological phase transition with gap closing. We find that the state transport exhibits universal power-law scaling behaviors with quenching rate, which enriches the Kibble-Zurek phenomena. More importantly,  starting from edge state or bulk state will produce distinct scaling exponents. After a periodic quench circle, Thouless pumping does not hold if the topological transition is crossed, but rather the returning probability to the initial edge state  is nonzero and exhibits also a power-law scaling with quenching rate. Our results are of broad interest in nonequilibrium quantum statistical mechanics, connecting quantum state transfer with the breakdown of adiabatic dynamics, and should find broad applications in quantum information, quantum annealing, ultracold atom physics, and the study of critical phenomena.

\acknowledgements{}
This paper was supported by the National Key Research
and Development Program of the Ministry of Science and Technology (Grant No. 2021YFA1200700), the National Natural Science Foundation of China (Grants No. 12275075, and No. 12105094), and the Fundamental Research Funds for the Central Universities of China.

\section*{Appendix A: Probabilities projected to extended eigenstates for linear quenches from an edge}

\setcounter{figure}{0}
\setcounter{equation}{0}
\renewcommand{\theequation}{A\arabic{equation}}
\renewcommand\thefigure{A\arabic{figure}}

In the main text, for linear quenches from the edge, our theoretical estimate of the travel distance is based on the picture that the edge state transfers to each extended state near $t=t_f/2$, which then travels at its group velocity. Here we justify this picture by calculating probabilities projected to each 
extended eigenstate.

We work on a chain with an odd number of sites, whose eigenvalues and eigenstates have analytical forms. The eigenvalues from lowest to highest (assuming that $J_1$ and $J_2$ have the same sign) are \cite{Coutant_2020}:
\begin{align}\label{eq:spectrum}
&\epsilon_j=-|J_1+J_2e^{-i k_j}|, \quad \textrm{for } j=1,2,\ldots ,N-1, \nn\\
&\epsilon_N=0,\nn\\
&\epsilon_j=|J_1+J_2e^{-i k_j}|, \quad \textrm{for } j=N+1,\ldots, 2N-1,
\end{align}
where $k_j=\pi j/N$ (Note that, unlike in the main text, we used the integer $j$ to label the states, which now to range from $1$ to $2N-1$. The definition of $\epsilon_j$ is also different from $\epsilon_k$; it can now take negative values.) The corresponding normalized eigenstates are:
\begin{align}\label{}
&|\psi_j\ra=\frac{1}{\sqrt{N}}\left[-\sum_{n=1}^N\sin(nk_j+\phi_j)|n,A\ra\right.\nn\\
&\left.+\sum_{n=1}^{N-1}\sin(n k_j)|n,B\ra\right],\quad  \textrm{for } j< N,
\end{align}
\begin{align}
&|\psi_N\ra=\sqrt{\frac{1-(J_1/J_2)^2}{1-(J_1/J_2)^{2N}}} \sum_{n=1}^N\left(-\frac{J_1}{J_2}\right)^{n-1}|n,A\ra,
\end{align}
\begin{align}\label{}
&|\psi_j\ra=\frac{1}{\sqrt{N}}\left[\sum_{n=1}^N\sin(nk_j+\phi_j)|n,A\ra\right.\nn\\
&\left.+\sum_{n=1}^{N-1}\sin(n k_j)|n,B\ra\right],\quad  \textrm{for } j> N,
\end{align}
where $\phi_j=\arg(J_1+J_2 e^{-i k_j})$ is a phase shift on sublattice $A$.  
$|\psi_N\ra$ is the edge state, which is localized at the left (right) end of the chain if $J_1<J_2$ ($J_1>J_2$). All other eigenstates are extended; 
note that 
they depend on time through $\phi_j$.

\begin{figure}[!htb]
\scalebox{0.5}[0.5]{\includegraphics{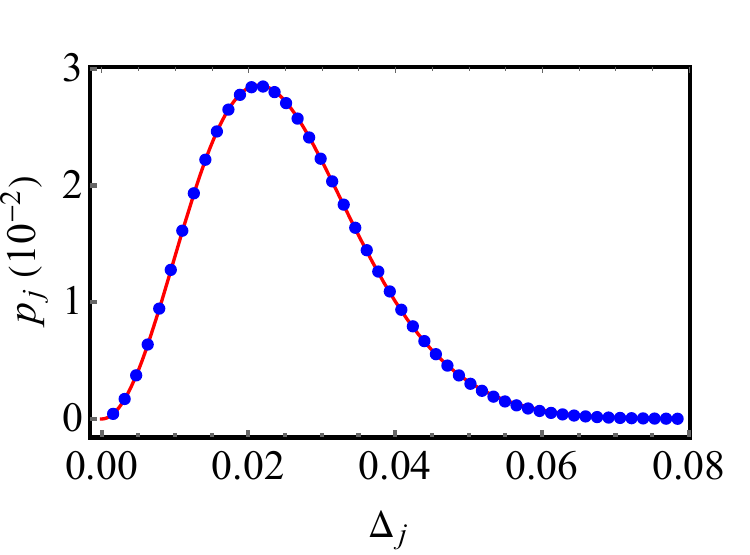}}
\caption{$p_{j}$ vs. $\Delta_{j}$ at $\beta=5\times 10^{-4}$ for $N=1000$ sites for evolutions from an edge. The blue dots are exact results, and the red line is a fitting by Eq.~\eqref{p_vs_epsilon_fit} with $c_1=1.03$, $c_2=0.657$ and $c_3 = 0.0856$.}
\label{fig:p_vs_epsilon}
\end{figure}

With these analytical expressions, one can readily calculate $|\la \psi_j|\psi(t)\ra|^2$, i.e. probabilities of the evolving state projected to each extended eigenstate at any time. 
We looked at $|\la \psi_j|\psi(t)\ra|^2$ at the final time $t_f$ of a linear quench, which we denote as $p_{j}$. We observe that $p_{j}=p_{2N-j}$, which originates from chiral symmetry, so it suffices to consider only the $j>N$ states. Fig.~\ref{fig:p_vs_epsilon} shows $p_{j}$ vs. $\Delta_{j}$ ($\Delta_{j}=|\cos(k_j/2)|$ is the gap between the $j$th extended state and the edge state at $t=t_f/2$) at $\beta=5\times10^{-4}$. We see that as $\Delta_{j}$ increases, $p_{j}$ increases from zero, reaches a peak and then decreases to zero. We find that for different $\beta$ the profiles of  $p_{j}$ vs. $\Delta_{j}$ can be fitted by a function of the form
\begin{align}\label{p_vs_epsilon_fit}
p_j=c_3 \left(1 - e^{-\frac{ c_1 \Delta_{j}^2}{\beta}}\right)e^{-\frac{ c_2 \Delta_{j}^2}{\beta}}.
\end{align}
The fitting parameters $c_1$ and $c_2$ depend slightly on $\beta$, and $c_3$ is an overall normalization factor. At $\beta=5\times10^{-4}$, $p_j$ reaches its maximum at $\Delta_j^2/\beta=0.967$. 

\begin{figure}[!htb]
\scalebox{0.5}[0.5]{\includegraphics{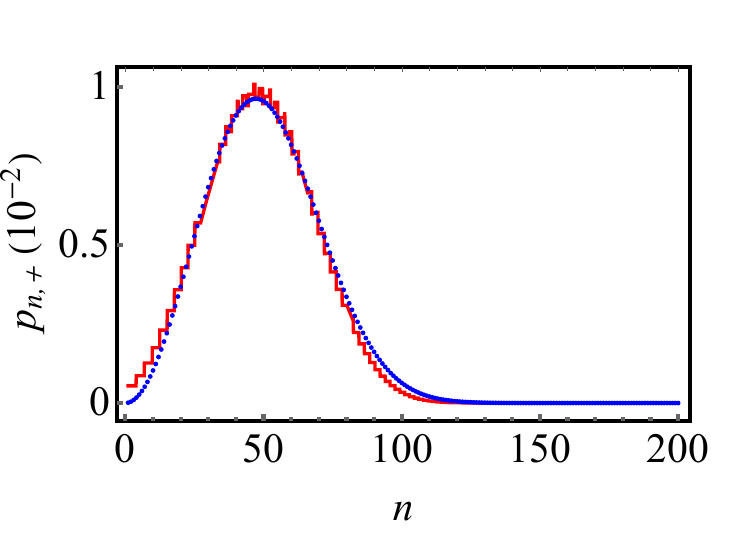}}
\caption{$p_{n,+}$ vs. $n$ at $\beta=5\times 10^{-4}$ for $N=1000$ sites for evolutions from an edge. The blue dots are exact results, and the red line is from the calculation using $p_{j}$ as described in the text, with each wave package taken as a rectangular function $(p_j/40) \operatorname{rect}[(n-1-d_j)/40]$.}
\label{fig:pn_from_pj}
\end{figure}

We further use  $p_{j}$ to calculate the final state's profile $p_{n,+}$, by assuming that the $j$th eigenstate travels at a speed $|v_k|=|\partial\epsilon_j/\partial k|$ and finally produces a wave package with a strength $p_j$ centered at $d_j=\int_{1/(2\beta)}^{1/\beta}|v_k| dt$, and the final profile $p_{n,+}$ is a summation of all these wave packages. Each wave package has a certain shape with a certain width which our theory does not predict, and we simply take it to be a rectangular function $(p_j/W) \operatorname{rect}[(n-1-d_j)/W]$, where the width $W$ (assumed to be the same for all $j$) is treated as a fitting parameter. Such a calculation gives a profile which reproduces well the exact $p_{n,+}$, as shown in Fig.~\ref{fig:pn_from_pj}. 
This further confirms the picture that the edge state transfers to each extended state and then travels at its group velocity.

\section*{Appendix B: Fidelity of adiabatic transfer}

\setcounter{figure}{0}
\setcounter{equation}{0}
\renewcommand{\theequation}{B\arabic{equation}}
\renewcommand\thefigure{B\arabic{figure}}

\begin{figure}[!htb]
\scalebox{0.5}[0.5]{\includegraphics{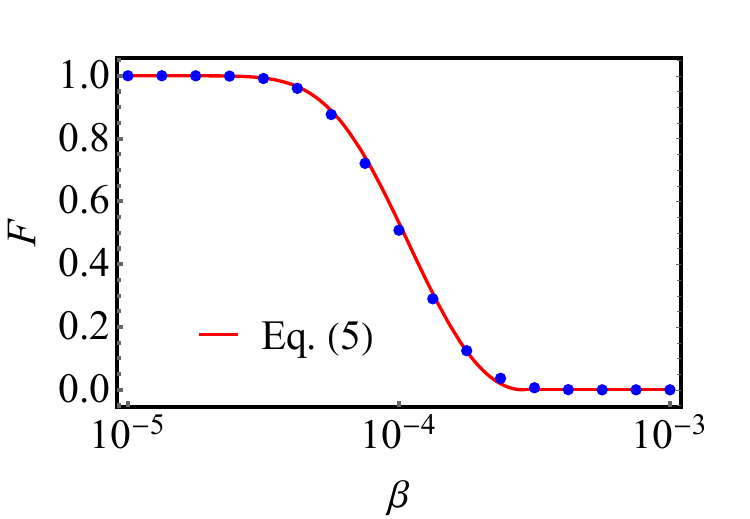}}
\caption{Fidelity of adiabatic transfer in the SSH model, $F$ vs. $\beta$, for $N=100$ for evolutions from an edge. The blue dots are exact results, and the red dashed line is Eq.~\eqref{eq:fidelity}.  }
\label{fig:fidelity}
\end{figure}

The power-law scalings we found take place when the state does not evolve to the other edge (the edges) of the chain. Namely, we have been effectively considering a semi-infinite (infinite) chain. For a finite chain with an odd number of sites, in the adiabatic limit $\beta\rar 0$ we expect that $|1,A\ra$ evolves adiabatically to $|N,A\ra$, since the edge state is always gapped from other states. 
For a larger $\beta$ this adiabatic transfer is not perfect, and its efficiency can be characterized by the fidelity $F=|\la N,B|\psi(t_f)\ra|^2$. In Fig.~\ref{fig:fidelity} we plot $F$ vs. $\beta$ for a chain with $N=100$ with an odd number of sites. We found that $F$ can be fitted by a simple formula:
\begin{align}\label{eq:fidelity}
F=\left[\max\left( 1-2e^{-\frac{2}{N^2\beta} },0\right)\right]^2,
\end{align}
which works for different values of $N$. The appearance of $1-2e^{-2/(N^2\beta)}$ in \eqref{eq:fidelity} reminds us the transition probability of staying on the middle level in the bow-tie model \cite{bow-tie}, and implies that the considered SSH model with linear time-dependence may be somehow similar to the bow-tie model.

\begin{thebibliography}{}
\bibitem{note-pk}
We found that transition probabilities from the edge state to the two extended states at the same $k$ with positive and negative energies are the same, which is due to chiral symmetry. They are thus denoted by the same symbol $p_k$.
\bibitem{note-LZ} 
One may think that a transition within a $2\times 2$ bulk, namely from $|\psi_{k,+}\ra$ to $|\psi_{k,-}\ra$ can take place, which may influence the estimate of $d_{bulk}$. Actually at the wavevector $k_{max}$ (which value to be determined shortly after) we will use to calculate $d_{T,bulk}$, such a transition can be ignored, due to the following reason. Each $2\times 2$ block can be described by a two-state LZ model $H=2\beta\sin(k/2)t\sigma_z+\cos(k/2) \sigma_x$, and the LZ model corresponding to the value of $k_{max}$ is in the adiabatic region since $\exp(-2\pi \Gamma)=\exp[-2\pi \cos^2(k_{max}/2)/(4\beta\sin(k_{max}/2))]\ll 1$ in the whole range of $\beta$ considered, so transition amplitudes between $|\psi_{k,+}\ra$ and $|\psi_{k,-}\ra$ are vanishingly small.
\bibitem{note-adiabatic} As argued in \cite{note-LZ}, when considering $p_{n,+}$ near the peak, evolutions of all $|k\ra\otimes|\psi_{k,+}\ra$ states involved can be treated as adiabatic.
\end{thebibliography}
\end{document}